\documentclass{sig-alternate-2013}
\usepackage{url}
\newcommand{\argmax}[1]{\underset{#1}{\operatorname{arg}\,\operatorname{max}}\;}

\usepackage{balance}
\usepackage{booktabs}
\usepackage{multirow}
\usepackage{ctable}
\usepackage{hyphsubst}
\usepackage{graphicx}
\usepackage{amsmath}
\graphicspath{{images/}}
\usepackage{times}
\usepackage{hhline}
\usepackage[caption = false]{subfig}
\permission{Permission to make digital or hard copies of all or part of this work for personal or classroom use is granted without fee provided that copies are not made or distributed for profit or commercial advantage and that copies bear this notice and the full citation on the first page. Copyrights for components of this work owned by others than ACM must be honored. Abstracting with credit is permitted. To copy otherwise, or republish, to post on servers or to redistribute to lists, requires prior specific permission and/or a fee. Request permissions from permissions@acm.org.}
\conferenceinfo{RecSys'14,}{Silicon Valley, USA, October 06-10, 2014.}
\copyrightetc{Copyright 2014 ACM \the\acmcopyr}
\crdata{xxx-x-xxxx-xxxx-x/xx/xx\ ...\$xx.xx}

\begin{document}
\title{Forgetting the Words but Remembering the Meaning:\\
Modeling Forgetting in a Verbal and Semantic Tag Recommender}

\numberofauthors{4} 
%
\author{
Dominik Kowald\\
       \affaddr{Know-Center}\\
       \affaddr{Graz University of Technology}\\
       \affaddr{Graz, Austria}\\
       \email{dkowald@know-center.at}
\alignauthor
Paul Seitlinger\\
       \affaddr{KTI}\\
       \affaddr{Graz University of Technology}\\
       \affaddr{Graz, Austria}\\
       \email{paul.seitlinger@tugraz.at}
       \and
Christoph Trattner\\
       \affaddr{Know-Center}\\
       \affaddr{Graz University of Technology}\\
       \affaddr{Graz, Austria}\\
       \email{ctrattner@know-center.at}
\alignauthor
Tobias Ley\\
       \affaddr{Institute of Informatics}\\
       \affaddr{Tallinn University}\\
       \affaddr{Tallinn, Estonia}\\
       \email{tley@tlu.ee}
}

\maketitle
\begin{abstract}
We assume that recommender systems are more successful, when they are based on a thorough understanding of how people process information. In the current paper we test this assumption in the context of social tagging systems. Cognitive research on how people assign tags has shown that they draw on two interconnected levels of knowledge in their memory: on a conceptual level of semantic fields or topics, and on a lexical level that turns patterns on the semantic level into words. Another strand of tagging research reveals a strong impact of time dependent forgetting on users' tag choices, such that recently used tags have a higher probability being reused than ``older'' tags. In this paper, we align both strands by implementing a computational theory of human memory that integrates the two-level conception and the process of forgetting in form of a tag recommender and test it in three large-scale social tagging datasets (drawn from BibSonomy, CiteULike and Flickr).

As expected, our results reveal a selective effect of time: forgetting is much more pronounced on the lexical level of tags. Second, an extensive evaluation based on this observation shows that a tag recommender interconnecting both levels and integrating time dependent forgetting on the lexical level results in high accuracy predictions and outperforms other well-established algorithms, such as Collaborative Filtering, Pairwise Interaction Tensor Factorization, FolkRank and two alternative time dependent approaches. We conclude that tag recommenders can benefit from going beyond the manifest level of word co-occurrences, and from including forgetting processes on the lexical level.

\end{abstract}


\\

\section{Introduction} \label{sec:intro}
%

Many interactive systems are designed in a way that they mimic human behavior and thinking. For example, intelligent tutoring systems make inferences similar to teachers when they draw on knowledge of the learning domain, knowledge about the learner and knowledge about effective teaching strategies. Similarly, recommender systems based on Collaborative Filtering use information about socially similar individuals to recommend items, much in the same way as humans are influenced by similar peers when they make choices. An implicit assumption behind this seems to be that interactive systems should be better the closer they correspond to human behavior. Such assumption seems to be sensible because it is humans that interact with these systems and the systems often draw on data that humans have produced (such as in the case of the Collaborative Filtering approaches). It is therefore reasonable to assume that strategies that have evolved in humans over their individual or collective development are good models for interactive systems. However, the assumption that an interactive system should perform better the closer it mimics human behavior is not often tested directly. 

\begin{figure*}[t!]
   \centering 
   \subfloat[BibSonomy]{ 
      \includegraphics[width=0.33\textwidth]{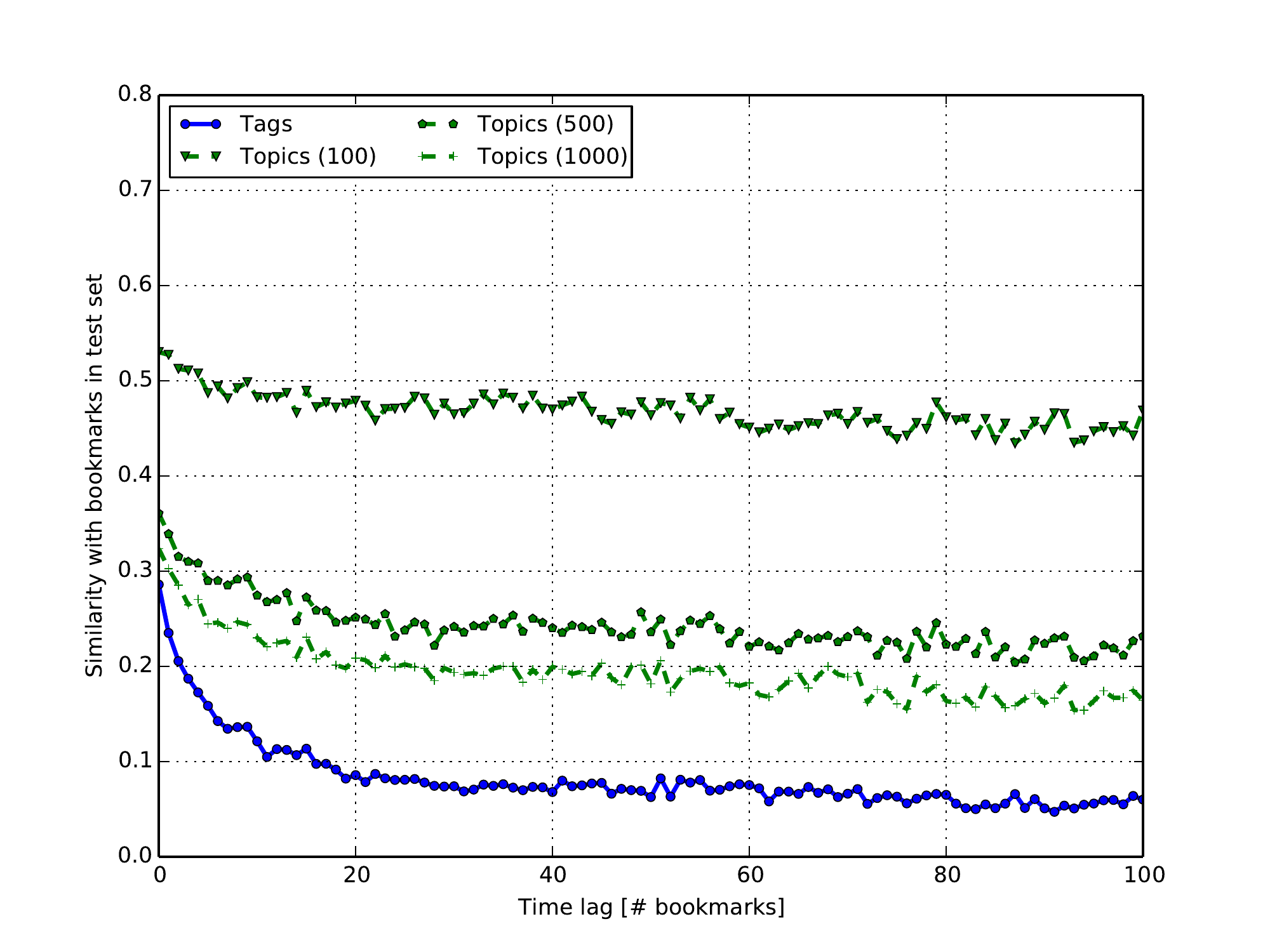}%
   } 
   \subfloat[CiteULike]{ 
      \includegraphics[width=0.33\textwidth]{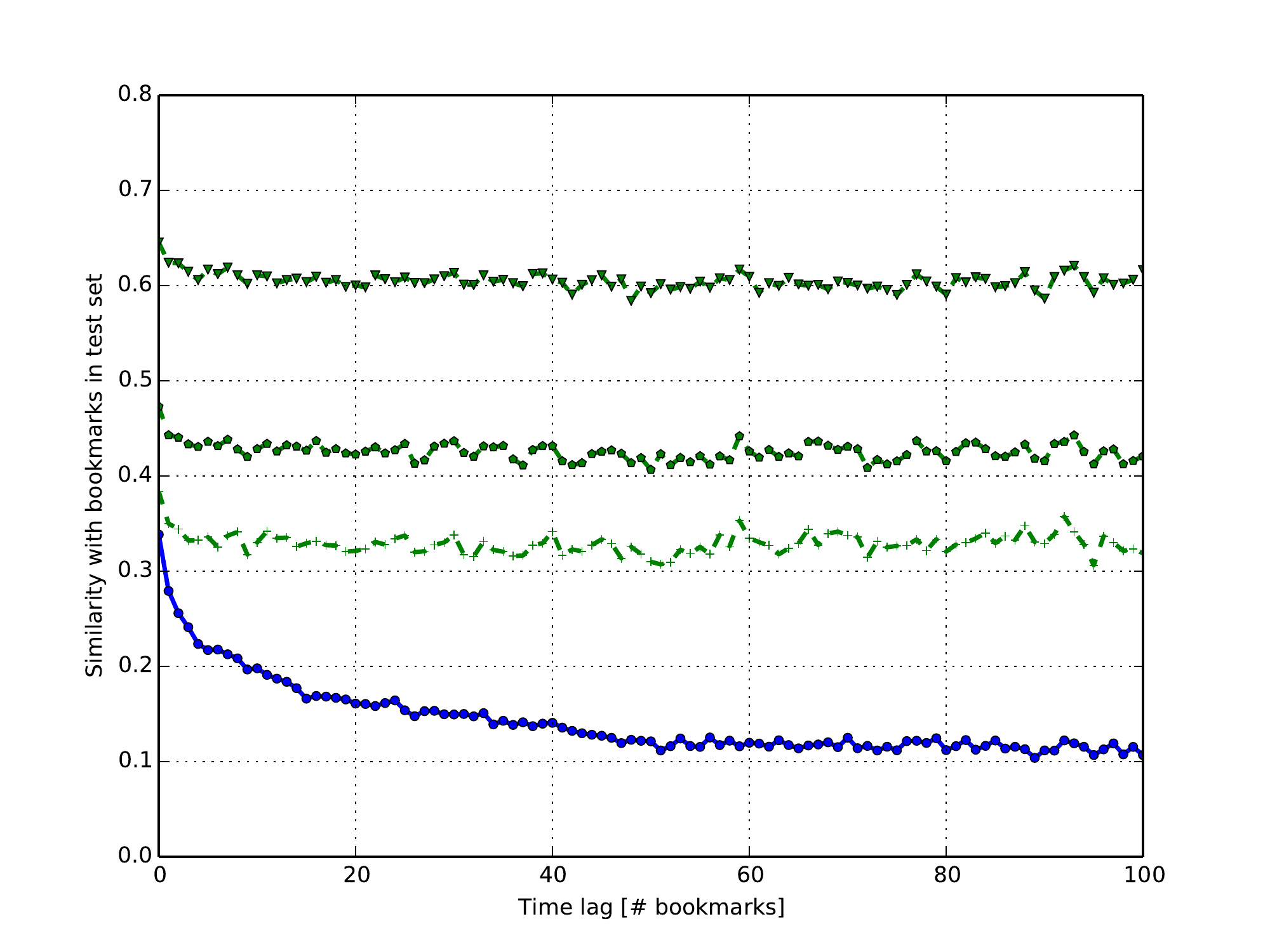}%
   } 
   \subfloat[Flickr]{ 
      \includegraphics[width=0.33\textwidth]{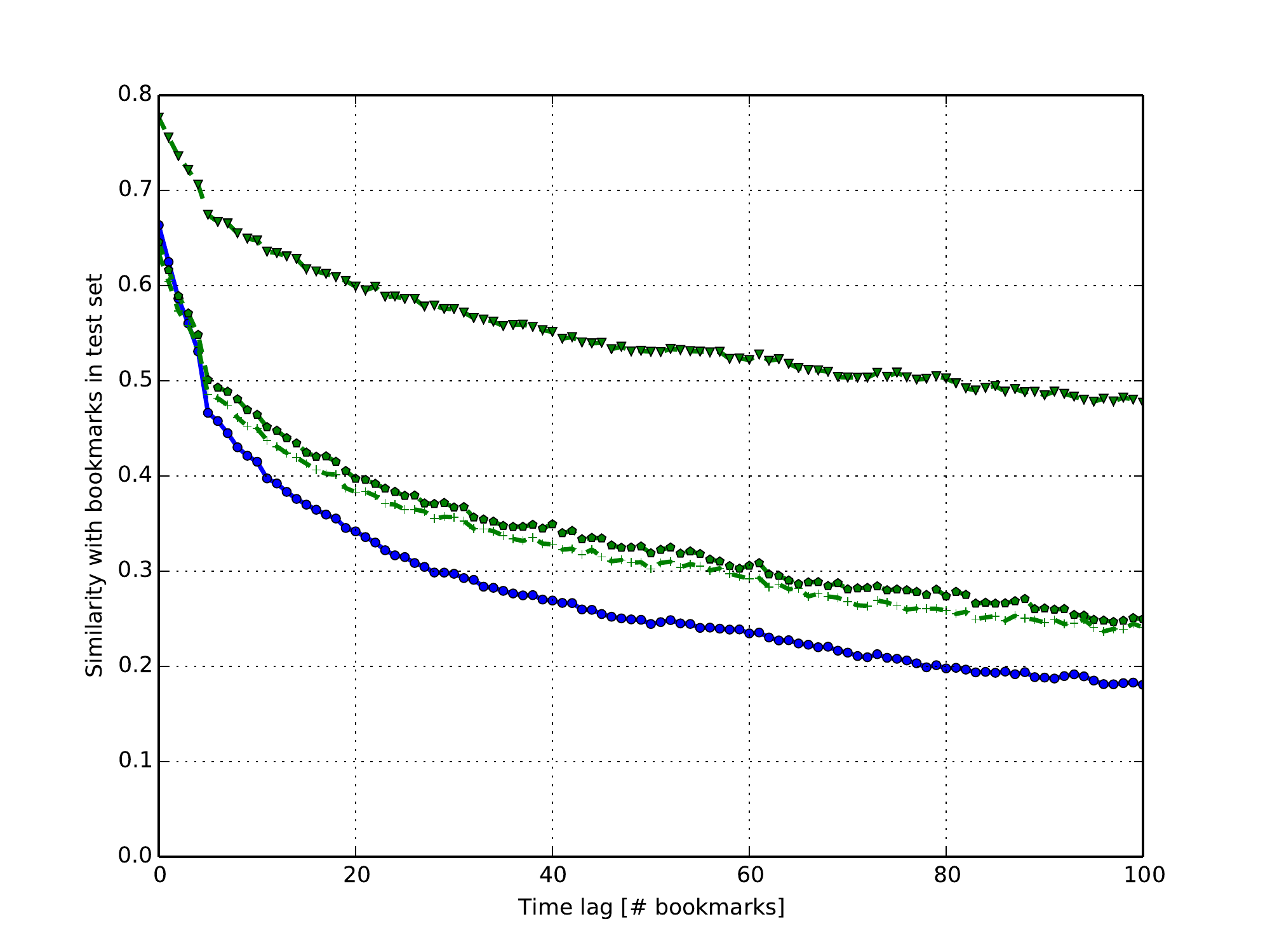}%
   } 
        \caption{Interaction between time dependent forgetting and level of knowledge representation for BibSonomy, CiteULike and Flickr showing a more pronounced decline for tags than for topics (100, 500, 1000 LDA topics) (first research question).}
				\label{fig:intro_plots}
\end{figure*}

In the current paper, we test this assumption in the context of a tag recommender algorithm. We draw on research that has explored how human memory is used in a dynamic and adaptive fashion to make sense of new information encountered in the environment. Sensemaking happens by dynamically forming ad-hoc categories that relate the new information with knowledge stored in the semantic memory (e.g., \cite{barsalou2003situated}). For instance, when reading an article about personalized recommendations, a novice has to figure out meaningful connections between previously distinct topics such as cognition and information retrieval and hence, has to start developing an ad-hoc category about common features of both of them. When using a social tagging system in such a situation, people apply labels to their own resources which to some extent externalize this process of spontaneously generating ad-hoc categories \cite{glushko2008categorization}. Usually, a user describes a particular bookmark by a combination of about three to five tags verbalizing and associating aspects of different topics (e.g., ``memory'',``retrieval'', ``recommendations'', ``collaborative filtering'').

In previous work, we have shown that this behavior can be well described by differentiating between two separate forms of information processing in human memory, a semantic process that generates and retrieves topics or gist traces, and a verbal process that generates verbatim word forms to describe the topics \cite{paul2013}. In this paper, we put an emphasis on another fundamental principle of human cognition to improve this model. According to Polyn et al. \cite{polyn2009context}, memory traces including recently activated features contribute more strongly to retrieval than traces including features that have not been activated for a longer period of time. This relationship provides a natural account of what is called the recency effect in memory psychology (e.g., \cite{anderson_reflections_1991}). Obviously, things that happened a longer time ago tend to be forgotten and influence our current behavior less than things that have happened recently.

The purpose of this paper is twofold. First, we study the interaction between the effect of recency and the level of knowledge representation in human memory (semantic vs. verbal) in a social tagging system. In particular, we raise the question whether the impact of recency interacts with the level of knowledge representation, i.e., whether a time-dependent shift in the use of topics can be dissociated from a time-dependent shift in the use of particular tags (\textit{first research question}). The second purpose, then, is to examine the question as to whether our tag recommender can be improved by integrating a time-dependent forgetting process and how this recommender performs in comparison to other well-established tag recommender algorithms (e.g., Collaborative Filtering, Pairwise Interaction Tensor Factorization and FolkRank), as well as two alternative time-dependent approach called GIRPTM \cite{zhang2012integrating} and BLL+C \cite{domi2014} (\textit{second research question}).


\begin{figure*}[t!]
        \centering
        \includegraphics[width=0.8\textwidth]{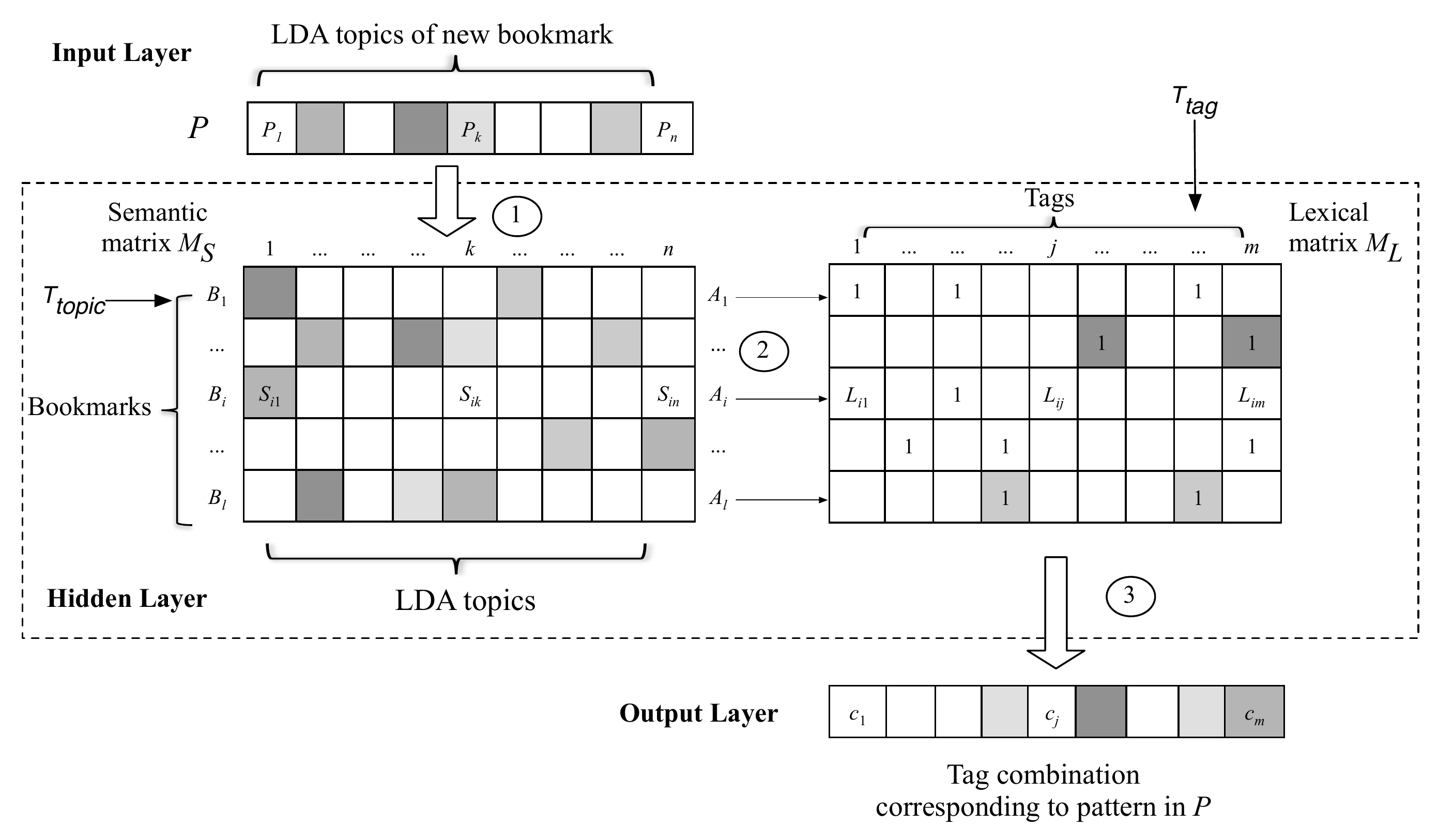}
        \caption{Schematic illustration of 3L showing the connections between the semantic matrix ($M_S$) encoding the LDA topics and the lexical matrix ($M_L$) encoding the tags. Furthermore, T$_{topic}$ and T$_{tag}$ schematically demonstrate how the time component is integrated in case of 3LT$_{topic}$ and 3LT$_{tag}$, respectively.} 
				\label{fig:approach}
\end{figure*}

The remainder of this paper is organized as follows. We begin with reviewing some of the work concerning recency in memory research and its current use in social tagging in Section \ref{sec:recency}. Then we describe our approach and the experimental setup of our extensive evaluation in Sections \ref{sec:approach} and \ref{sec:exp_setup}. We then present the results of this evaluation in terms of recommender quality in Section \ref{sec:results} and discuss related work in the field in Section \ref{sec:relwork}. Finally, we conclude the paper by discussing our findings and future work in Section \ref{sec:disc_con}.

\section{Recency in Memory and in the Use of Social Tagging} \label{sec:recency}
In previous work we have introduced 3Layers \cite{paul2013}, a model for recommending tags that is inspired by cognitive-psychological research on categorizing and verbalizing objects (e.g., \cite{glushko2008categorization}). It consists of an input, a hidden and an output layer, where the hidden layer is built up by a semantic and an interconnected lexical matrix.
The semantic matrix stores the topics of all bookmarks in the user's personomy, calculated with Latent Dirichlet Allocation (LDA) \cite{krestel2009latent}, while the lexical matrix stores the tags of those bookmarks. In a first step of calculation, the LDA topics of a new bookmark, for which appropriate tags should be recommended, are represented in the input layer and compared with the semantic matrix of the hidden layer. In the course of this comparison, semantically relevant bookmarks of the user's personomy become activated. 
The resulting pattern of activation across the semantic matrix is then applied to the lexical matrix to further activate and recommend those tags that belong to relevant bookmarks. 
In a final step, the activation pattern across the lexical matrix is summarized on the output layer in form of a vector representing a tag distribution that can be used to predict a substantial amount of variance in the user's tagging behavior for the new bookmark.

We draw on Fuzzy Trace Theory (FTT; e.g., \cite{brainerd2010recollective}) to make a prediction with respect to our first research question about a potentially differential impact of recency on semantic and lexical representations, i.e., on the usage of topics and tags, respectively.
FTT differentiates between two distinct memory traces, a gist trace and a verbatim trace, which represent general semantic information of e.g., a read sentence and the sentence's exact wording, respectively. These two types of memory traces share properties with our distinction between a semantic and a lexical matrix (see also Section \ref{sec:approach}). While vectors of the semantic matrix provide a formal account of each bookmark's gist (its general semantic content), vectors of the lexical matrix correspond to a bookmark's verbatim trace (explicit verbal information in form of assigned tags).

This assumption is also in line with Kintsch \& Mangalath \cite{kintsch2011construction} who model gist traces of words by means of LDA topic vectors and explicit traces of words by means of word co-occurrence vectors. An empirically well-established assumption of FTT is that verbatim traces are much more prone to time-dependent forgetting than gist traces (e.g., \cite{brainerd2010recollective}): while people tend to forget the exact wording, usually they can remember the gist of a sentence (or a bookmark). Taken together, we derived the hypothesis that a user's verbatim traces (vectors in the lexical matrix that encode the user's tags) are more strongly affected by time-dependent forgetting and therefore more variable over time than a user's gist traces (vectors in the semantic matrix) that should be more similar to each other over time.

To test this hypothesis, we performed an empirical analysis in BibSonomy, CiteULike and Flickr (see Section \ref{sec:datasets}). The topics for the resources of these datasets' bookmarks were calculated using Latent Dirichlet Allocation (LDA) \cite{krestel2009latent} (see Section \ref{sec:approach}) based on 100, 500 and 1000 latent topics in order to cover different levels of topic specialization. For each user we selected the most recent bookmark (i.e., the one from the test set with the largest timestamp, see also Section \ref{sec:evaluation_metrics}) and described the bookmark by means of two vectors: one encoding the bookmark's LDA topic pattern (gist vector) and one encoding the tags assigned by the user (verbatim vector). Then, we searched for all the remaining bookmarks of the same user, described each of them by means of the two vectors and arranged them in a chronologically descending order. Next, we compared the gist and the verbatim vector of the most recent bookmark with the two corresponding vectors of all bookmarks in the user's past by means of the cosine similarity measure.

The obtained results are represented in the three diagrams of Figure \ref{fig:intro_plots}, plotting the average cosine similarities over all users against the time lags in days. For all three datasets we show these results for the last 100 bookmarks of tagging activity because in this range there are enough users available for each bookmark to calculate mean values reliably. The diagrams quite clearly reveal that -- independent of the environment (BibSonomy, CiteULike or Flickr) -- the similarity between the most recent bookmark and all other bookmarks decreases monotonically as a function of time lag. More importantly and as expected, the time dependent decline is more strongly pronounced for the verbatim vectors (encoding tag assignments) in contrast to the gist vectors (encoding LDA topics). Furthermore, we can seen that the more LDA topics we use, the more similar is the time dependent decline of the two vectors (tags vs. topics) to each other. 

\section{Approach} \label{sec:approach}
In this section we introduce two novel time-dependent tag recommender algorithms which model the process of forgetting on a semantic and lexical layer in a time-depended manner. Based on our findings from the previous section, we assume that the factor of time plays a more critical role on the lexical layer than on the semantic layer. The approaches implemented in this section are based on a preliminary concept called 3Layers that was introduced in our previous work \cite{paul2013} to model semantic and lexical processes of tagging in social bookmarking systems.

Figure \ref{fig:approach} schematically shows how 3Layers (3L) represents a user's personomy within the hidden layer, which interconnects a semantic matrix, $M_S$ ($l$ bookmarks $\times$ $n$ LDA topics matrix), and a lexical matrix, $M_L$ ($l$ bookmarks $\times$ $m$ tags matrix). Thus, each bookmark of the user is represented by two associated vectors; by a vector of LDA topics $S_{i,k}$ stored in $M_S$ and by a vector of tags $L_{i,j}$ stored in $M_L$. Similar to \cite{kwantes2005using}, we borrow a mechanism from MINERVA2, a computational theory of human categorization \cite{hintzman1984minerva}, to process the network constituted by the input, hidden and output layer. First, the LDA topics of the new resource to be tagged are represented on the input layer in form of a vector $P$ with $n$ features. Then, $P$ is used as a cue to activate each bookmark ($B_i$) in $M_S$ depending on the similarity ($Sim_i$) between both vectors, i.e., $P$ and $B_i$. Similar to \cite{kwantes2005using}, we estimate $Sim_i$ by calculating the cosine between the two vectors:
	\begin{align}
	Sim_{i} = \frac{\sum_{k=1}^{n}P_{k} \times S_{i,k}}{\sqrt{\sum_{k=1}^{n}P_{k}^{2}} \times \sqrt{\sum_{k=1}^{n}S_{i,k}^{2}}}
	\end{align}
To transform the resulting similarity values into activation values ($A_i$) and to further reduce the influence of bookmarks with low similarities, $Sim_i$ is raised to the power of 3, i.e. $A_i = Sim_i^3$ (see also \cite{hintzman1984minerva}). Next, these activation values are propagated to $M_L$ to activate tags that are associated with highly activated bookmarks on the semantic matrix $M_S$ (circled numbers 2 and 3 in Figure \ref{fig:approach}). This is realized by the following equation that yields an activation value $c_j$ for each of the $m$ tags on the output layer:
	\begin{align}
	\underbrace{c_{j} = \sum_{i=1}^{l}(L_{i,j} \times A_{i})}_{3L}
	\end{align}
To finally realize 3LT$_{topic}$ and 3LT$_{tag}$, we integrate a time component on the level of topics (hereinafter called T$_{topic}$) and on the level of tags (T$_{tag}$), respectively. Both recency components are calculated by the following equation that is based on the base-level learning (BLL) equation \cite{anderson_reflections_1991}:
	\begin{align}
	BLL(t) = ln((tmstp_{ref} - tmstp_{t})^{-d})
	\end{align}
, where $tmstp_{ref}$ is the timestamp of the most recent bookmark of the user and $tmstp_{t}$ is the timestamp of the last occurrence of \textit{t}, encoded as the topic in the case of T$_{topic}$ or as the tag in the case of T$_{tag}$, in the user's bookmarks. The exponent $d$ accounts for the power-law of forgetting and was set to $0.5$ as suggested by Anderson et al. \cite{Anderson2004}.
While 3LT$_{topic}$ can be realized by using equation (4), 3LT$_{tag}$ can be realized by using equation (5):
	\begin{align}
	\underbrace{c_{j} = \sum_{i=1}^{l}(L_{i,j} \times \sum_{k=1}^{n}(S_{i,k} \times BLL(k)) \times A_{i})}_{3LT_{topic}}
	\end{align}
	\begin{align}
	\underbrace{c_{j} = \sum_{i=1}^{l}(L_{i,j} \times BLL(j) \times A_{i})}_{3LT_{tag}}
	\end{align}

As described in \cite{domi2014}, it is also important to take into account the tags that have been applied by other users to the target resource in the past in order to be able to also recommend new tags, i.e. tags that have not been used by the target user before. We do this by simply taking into account the most popular tags in the tag assignments of the resource Y$_{r}$ (MP$_r$, i.e., $\argmax{t \in T}(|Y_{r}|)$) \cite{hotho2006information}. In order to combine $c_j$ with MP$_r$, the following normalization method was used:
	\begin{align}
	\|c_{j}\| = \frac{exp(c_{j})}{\sum_{i=1}^{m}exp(c_{i})}
	\end{align}

Taken together, the list of recommended tags for a given user $u$ and resource $r$ is then calculated as
  \begin{align}
		\widetilde{T}(u, r) = \argmax{j \in T}(\beta \|c_{j}\| + (1 - \beta) \||Y_{r}|\|)
  \end{align}
, where $\beta$ is used to inversely weight the two components. The results presented in Section \ref{sec:results} were calculated using $\beta$ = 0.5, so giving the same weight to both components. The algorithms presented in this work are implemented in the Java programming language, are open-source software and can be exported online from our Github Repository\footnote{\url{https://github.com/learning-layers/TagRec/}} along with the test and training sets used for our experiments (see Section \ref{sec:datasets} and \ref{sec:evaluation_metrics}).

As outlined in Section \ref{sec:recency}, we calculated the semantic features of the resources in the bookmarks using \textit{Latent Dirichlet Allocation} (LDA). In general, LDA is a probability model that helps to find latent topics for documents where each topic is described by words in these documents \cite{krestel2009latent}. This can be formalized as follows:
\begin{align}
		P(t_{i}|d) = \sum\limits_{ j=1 }\limits^{ Z }{ P(t_{i} | z_{i} = j) P(z_{i} = j | d)}
  \end{align}
Here $P(t_i|d)$ is the probability of the $i$th word for a document $d$ and $P(t_i|z_i = j)$ is the probability of $t_i$ within the topic $z_i$. $P(z_i = j|d)$ is the probability of using a word from topic $z_i$ in the document. The number of latent topics $Z$ has to be chosen in advance, which defines the level of specialization of the topics. We calculated the semantic features for our datasets based on different numbers of LDA topics (100, 500 and 1000 - see also Section \ref{sec:results}).


When using LDA in tagging environments, documents are resources which are described by tags. This means that resources in the bookmarks can also be represented with the topics identified by LDA based on the tag vectors of the resources (i.e., all the tags the users have assigned to the resource). These topics were then used as features in the semantic matrix $M_S$. We implemented LDA with Gibbs sampling using the Java framework Mallet\footnote{\url{http://mallet.cs.umass.edu/topics.php}}.

\section{Experimental Setup} \label{sec:exp_setup}
In this section we describe in detail the datasets, the evaluation methodology and the baseline algorithms used for our experiments.

\begin{table}[t!]
\small
  \setlength{\tabcolsep}{2.6pt}
  \centering
    \begin{tabular}{l|lllll}
    \specialrule{.2em}{.1em}{.1em}
      Dataset			& $|B|$			& $|U|$	& $|R|$	& $|T|$	& $|TAS|$	 							\\\hline 
      BibSonomy	  & 400,983 & 5,488  & 346,444		& 103,503	& 1,479,970				\\\hline
			CiteULike		& 379,068 & 8,322  & 352,343		& 138,091	& 1,751,347				\\\hline															
			Flickr			& 864,679 & 9,590  & 864,679		& 127,599	& 3,552,540				\\ 
									
		\specialrule{.2em}{.1em}{.1em}								
    \end{tabular}
    \caption{Properties of the datasets, where $|B|$ is the number of bookmarks, $|U|$ the number of users, $|R|$ the number of resources, $|T|$ the number of tags and $|TAS|$ the number of tag assignments.}
  \label{tab:dataset_stats}
\end{table}

\subsection{Datasets} \label{sec:datasets}
We used three well-known folksonomy datasets that are freely available for scientific purposes in order to conduct our study and for reasons of reproducibility. In this respect, we utilized datasets from the social bookmark and publication sharing system BibSonomy\footnote{\url{http://www.kde.cs.uni-kassel.de/bibsonomy/dumps/}} (2013-07-01), the reference management system CiteULike\footnote{\url{http://www.citeulike.org/faq/data.adp}} (2013-03-10) and the image sharing platform Flickr\footnote{\url{http://www.tagora-project.eu/data/#flickrphotos}} (2010-01-07) to evaluate our approach on both types of folksonomies, broad (BibSonomy and CiteULike; all users are allowed to annotate a particular resource) and narrow (Flickr; only the user who has uploaded a resource is allowed to tag it) ones \cite{Helic2012}.
We furthermore excluded all automatically generated tags from the datasets (e.g., \textit{no-tag}, \textit{bibtex-import}, etc.) and decapitalized all tags as suggested by related work in the field (e.g., \cite{Rendle2010}). In the case of CiteULike we randomly selected 10\% and in the case of Flickr 3\% of the user profiles for reasons of computational effort (see also \cite{gemmell2009improving})\footnote{\textbf{Note:} We used the same dataset samples as in our previous work \cite{domi2014}, except of CiteULike, where we used a smaller sample for reasons of computational effort in respect to the calculation of the LDA topics.}. A $p$-core pruning approach was not applied in order to capture also the issue of cold-start users or items and to prevent a biased evaluation \cite{Doerfel2013}. The statistics of our datasets can be found in Table \ref{tab:dataset_stats}. 


\subsection{Evaluation Methodology} \label{sec:evaluation_metrics}
To evaluate our tag recommender approaches, we split the three datasets into training and test sets based on a leave-one-out hold-out method as proposed by related work in this field (e.g., \cite{jaschke2007tag}). Hence, for each user we selected her most recent bookmark (in time) and put it into the test set. The remaining bookmarks were then used for the training of the algorithms. This procedure simulates well a real-world environment because the tagging behavior of a user in the future is tried to be predicted based on the tagging behavior in the past. Furthermore, it is a standard procedure for the evaluation of time-based recommender systems \cite{campos2013time}. 

To finally quantify the recommender quality and to benchmark our recommender against other tag recommendation approaches, a set of well-known metrics in information retrieval and recommender systems were used. In particular, we report Recall (R@$k$), Precision (P@$k$), Mean Reciprocal Rank (MRR) and Mean Average Precision (MAP) for $k$ = 10 and F1-Score ($F_1$@$k$) for $k$ = 5 recommended tags\footnote{$F_1$@$5$ was also used as the main performance metric in the ECML PKDD Discovery Challenge 2009: http://www.kde.cs.uni-kassel.de/ws/dc09/.} \cite{lipczak2012hybrid}.

\subsection{Baseline Algorithms}\label{sec:algos}
We compared the results of our approach to several ``baseline'' tag recommender algorithms. The algorithms were selected in respect to their popularity in the community, performance and novelty \cite{balbymarinho2012recommender}.
The most basic approach we utilized is the unpersonalized \textit{MostPopular (MP)} algorithm that recommends for any user and any resource the same set of tags that is weighted by the frequency in all tag assignments \cite{jaschke2007tag}. A personalized extension of MP is the  \textit{MostPopular$_{u, r}$ (MP$_{u,r}$)} algorithm the suggests the most frequent tags in the tag assignments of the user (\textit{MP$_{u}$}) and the resource (\textit{MP$_{r}$}) \cite{jaschke2007tag}. Another simple and classic recommender approach is \textit{Collaborative Filtering (CF)} which was adapted for tag recommendations by Marinho et al. \cite{marinho2008collaborative}. Here the neighborhood of an user is formed based on the tag assignments in the user profile and the only variable parameter is the number of users $k$ in this neighborhood. $k$ has been set to 20 based on the work of Gemmell et al. \cite{gemmell2009improving}.

Another approach we utilized is the well-known \textit{FolkRank (FR)} algorithm which is an improvement of the \textit{Adapted PageRank (APR)} approach \cite{jaschke2007tag}. 
FR adapts the PageRank algorithm in order to rank the nodes within the graph structure of a folksonomy\cite{jaschke2007tag} based on their importance in the network. Our implementation of APR and FR is based on the code and the settings of the open-source Java tag recommender framework provided by the University of Kassel\footnote{\url{http://www.kde.cs.uni-kassel.de/code}}. A different popular and recent tag recommender mechanism is \textit{Pairwise Interaction Tensor Factorization (PITF)} proposed by Rendle \& Schmidt-Thieme \cite{Rendle2010}. It is an extension of \textit{Factorization Machines (FM)} and explicitly models the pairwise interactions between users, resources and tags. The FM and PITF results presented in this paper were calculated using the open-source C++ tag recommender framework provided by the University of Konstanz\footnote{\url{http://www.informatik.uni-konstanz.de/rendle/software/tag-recommender/}} with 256 factors as suggested in \cite{Rendle2010}.

Finally, we also tried to benchmark against two time dependent approaches. The first one is the \textit{GIRPTM} algorithm presented by Zhang et al. \cite{zhang2012integrating} which is based on the frequency and the temporal usage of a user's tag assignments. The approach models the temporal tag usage with an exponential distribution based on the first- and last-time usage of the tags. The second time-dependent tag-recommender approach is the \textit{Base-Level Learning Equation with Context (BLL+C)} algorithm introduced in our previous work \cite{domi2014}. BLL+C is based on the ACT-R human memory theory by Anderson et al. \cite{Anderson2004} and uses a power-law distribution based on all tag usages to mimic the time dependent forgetting in tag applications. In both approaches the resource component is modeled by a simple most popular tags by resource mechanism, as it is also done in our 3Layers approaches. 

\begin{table}[t!]
  \setlength{\tabcolsep}{3pt}
  \centering
  \small
    \begin{tabular}{c|l|l|lll}
    \specialrule{.2em}{.1em}{.1em}
			& \# Topics	& Measure 		& 3L		& 3LT$_{topic}$ 	 	& 3LT$_{tag}$ 								\\\hline
			\multirow{9}{*}{\centering{\begin{sideways}\centering{BibSonomy}\end{sideways}}}
			&	100				& $F_1$@5				&	.197			&	.198	 						& \textbf{.204}				\\
			&						&	MRR						&	.152			&	.154	 						& \textbf{.161}				\\
			&						&	MAP						&	.201			&	.202	 						& \textbf{.212}				\\
			&	500				& $F_1$@5				&	.204			&	.205	 						& \textbf{.209}				\\
			&						&	MRR						&	.156			&	.158	 						& \textbf{.163}				\\
			&						&	MAP						&	.206			&	.208	 						& \textbf{.215}				\\
			&	1000			& $F_1$@5				&	.206			&	.207	 						& \textbf{.211}				\\
			&						&	MRR						&	.157			&	.158	 						& \textbf{.162}				\\
			&						&	MAP						&	.207			&	.208	 						& \textbf{.214}				\\\hline				
			
			\multirow{9}{*}{\centering{\begin{sideways}\centering{CiteULike}\end{sideways}}}        				
			&	100				& $F_1$@5			&	.211			&	.212	 						& \textbf{.221}				\\
			&						&	MRR					&	.192			&	.194	 						& \textbf{.211}				\\
			&						&	MAP					&	.226			&	.228	 						& \textbf{.248}				\\
			&	500				& $F_1$@5			&	.218			&	.219	 						& \textbf{.225}				\\
			&						&	MRR					&	.196			&	.198	 						& \textbf{.211}				\\
			&						&	MAP					&	.232			&	.234	 						& \textbf{.250}				\\
			&	1000			& $F_1$@5			&	.232			&	.233	 						& \textbf{.238}				\\
			&						&	MRR					&	.199			&	.200	 						& \textbf{.212}				\\
			&						&	MAP					&	.235			&	.236	 						& \textbf{.250}				\\\hline
																																																												
			\multirow{9}{*}{\centering{\begin{sideways}\centering{Flickr}\end{sideways}}}		  																																		
			&	100				& $F_1$@5			&	.500			&	.507	 						& \textbf{.535}				\\
			&						&	MRR					&	.421			&	.429	 						& \textbf{.476}				\\
			&						&	MAP					&	.560			&	.571	 						& \textbf{.634}				\\
			&	500				& $F_1$@5			&	.564			&	.567	 						& \textbf{.582}				\\
			&						&	MRR					&	.443			&	.448	 						& \textbf{.476}				\\
			&						&	MAP					&	.591			&	.596	 						& \textbf{.635}				\\
			&	1000			& $F_1$@5			&	.568			&	.571	 						& \textbf{.585}				\\
			&						&	MRR					&	.450			&	.454	 						& \textbf{.477}				\\
			&						&	MAP					&	.599			&	.604	 						& \textbf{.636}				\\				
		\specialrule{.2em}{.1em}{.1em}		
    \end{tabular}
    \caption{$F_1$@5, MRR and MAP values for BibSonomy, CiteULike and Flickr showing the performance of 3L and its time dependent extensions (3LT$_{topic}$ and 3LT$_{tag}$) for 100, 500 and 1000 LDA topics (first research question).} 
  \label{tab:mrr_map_norm_layers}
\end{table}

\section{Results} \label{sec:results}

\begin{table*}[t!]
  \setlength{\tabcolsep}{3pt}
  \centering
  \small
    \begin{tabular}{c|l|l|lllllllllll|llll}
    \specialrule{.2em}{.1em}{.1em}
			& |$B_{min}$|			& Measure & MP		& LDA		& MP$_{u}$	& MP$_{r}$		& MP$_{u,r}$	 	& CF 			& APR		& FR 			& FM		& PITF		& GIRPTM  & BLL+C				& 3L		& 3LT$_{topic}$ 	 	& 3LT$_{tag}$ 				\\ \hline
			\multirow{6}{*}{\centering{\begin{sideways}\centering{BibSonomy}\end{sideways}}}
			&	-					& $F_1$@5	& .013 	& .097	&	.152			&	.074				& .192					& .166	 	& .175	& .171		& .122	&	.139		& .197	  &	.201							&	.206	&	.207	 						& \textbf{.211$^{\substack{\phantom{1}\\\phantom{1}}}$}			  \\
			&						&	MRR			& .008  & .083	&	.114			&	.054				& .148					& .133	 	& .149	& .148		& .097	&	.120		& .152	  &	.158							&	.157	&	.158	 						& \textbf{.162$^{\substack{\phantom{1}\\\phantom{1}}}$}			  \\
			&						&	MAP			& .009  & .101	&	.148			&	.070				& .194					& .173	 	& .193	& .194		& .120	&	.150		& .200	  &	.207							&	.207	&	.208	 						& \textbf{.214$^{\substack{\phantom{1}\\\phantom{1}}}$}			  \\\cline{2-18} 	
							
			&	20				& $F_1$@5	& .019 	& .142	&	.156			&	.078				& .195					& .204	 	& .184	& .197		& .162	&	.163		& .240	  &	.249							&	.264	&	.269	 						& \textbf{.296$^{\substack{**\\\circ}}$}				\\
			&						&	MRR			& .011  & .129	&	.135			&	.059				& .160					& .175	 	& .159	& .171		& .135	&	.137		& .201	  &	.216							&	.224	&	.227	 						& \textbf{.251$^{\substack{**\\\phantom{1}}}$}				\\
			&						&	MAP			& .012  & .152	&	.163			&	.074				& .200					& .219	 	& .197	& .214		& .164	&	.166		& .256	  &	.275							&	.289	&	.291	 						& \textbf{.325$^{\substack{**\\\circ}}$}	\\\hline 				
						
			\multirow{6}{*}{\centering{\begin{sideways}\centering{CiteULike}\end{sideways}}}        							
			&	-					& $F_1$@5	& .007 	& .068	&	.182			&	.033				& .199					& .157	 	& .162	& .160		& .113	&	.130		& .207	  &	.215							&	.232	&	.233	 						& \textbf{.238$^{\substack{**\\\phantom{1}}}$}					\\
			&						&	MRR			& .005  & .065	&	.164			&	.024				& .179					& .168	 	& .181	& .181		& .116	&	.149		& .196	  &	.205							&	.199	&	.200	 						& \textbf{.212$^{\substack{\phantom{1}\\\phantom{1}}}$}					\\
			&						&	MAP			& .005  & .073	&	.191			&	.029				& .210					& .196	 	& .212	& .212		& .132	&	.169		& .229	  &	.241							&	.235	&	.236	 						& \textbf{.250$^{\substack{\phantom{1}\\\phantom{1}}}$}					\\\cline{2-18}
							
			&	20				& $F_1$@5	& .008 	& .145	&	.228			&	.031				& .237					& .228	 	& .221	& .225		& .193	&	.196		& .282	  &	.298							&	.331$^{*}$	&	.334$^{*}$	& \textbf{.353$^{\substack{***\\\circ}}$}				\\
			&						&	MRR			& .006  & .144	&	.225			&	.022				& .233					& .271	 	& .237	& .239		& .201	&	.210		& .321	  &	.335							&	.312	&	.316	 						& \textbf{.367}$^{\substack{**\\\circ\circ\circ}}$				\\
			&						&	MAP			& .006  & .162	&	.258			&	.028				& .269					& .308	 	& .273	& .276		& .229	&	.237		& .369	  &	.389							&	.369	&	.373	 						& \textbf{.430}$^{\substack{**\\\circ\circ\circ}}$				\\\hline 	
																																																															
			\multirow{6}{*}{\centering{\begin{sideways}\centering{Flickr}\end{sideways}}}		  																																					
			&	-					& $F_1$@5	& .023 	& .169	&	.435			& -						& .435					& .417		& .328 	& .334		&	.297	&	.316		& .509		& .523							&	.568$^{***}$	&	.571$^{***}$	 	& \textbf{.585$^{\substack{***\\\circ}}$}		\\
			&						&	MRR			& .023  & .171	&	.360			& -						& .360					& .436		& .352	& .355		&	.300	&	.333		& .445		& .466							&	.450	&	.454	 			& \textbf{.477$^{\substack{*\\\circ\circ\circ}}$}			\\
			&						&	MAP			& .023  & .205	&	.468			& -						& .468					& .581		& .453 	& .459		&	.384	&	.426		& .590		& .619							&	.599	&	.604	 	& \textbf{.636$^{\substack{*\\\circ\circ\circ}}$}			\\\cline{2-18}
			
			&	20		 		& $F_1$@5	& .030 	& .190	&	.382			&	-						& .382					& .495	 	& .322	& .334		& .309	&	.309		& .534	  &	.553							&	.610$^{***}$	&	.616$^{***}$	 	& \textbf{.643$^{\substack{***\\\circ\circ\circ}}$}				\\
			&						&	MRR			& .028  & .174	&	.322			&	-						& .322					& .473	 	& .309	& .317		& .290	&	.289		& .485	  &	.508							&	.478	&	.485	 	& \textbf{.530}$^{\substack{**\\\circ\circ\circ}}$				\\
			&						&	MAP			& .029  & .215	&	.427			&	-						& .427					& .655	 	& .405	& .419		& .378	&	.376		& .664	  &	.701							&	.661	&	.670	 	& \textbf{.732}$^{\substack{***\\\circ\circ\circ}}$				\\
		\specialrule{.2em}{.1em}{.1em}		
    \end{tabular}
    \caption{$F_1$@5, MRR and MAP values for all the users in the datasets (BibSonomy, CiteULike and Flickr) and for users with a minimum number of 20 bookmarks (|$B_{min}$| = 20) showing that our time dependent 3LT$_{tag}$ approach outperforms current state-of-the art algorithms (second research question). The symbols $^{*}$, $^{**}$ and $^{***}$ indicate statistically significant differences based on a Wilcoxon Ranked Sum test between 3L, 3LT$_{topic}$, 3LT$_{tag}$ and BLL+C at $\alpha$ level .05, .01 and .001, respectively; $^{\circ}$, $^{\circ\circ}$ and $^{\circ\circ\circ}$ indicate statistically significant differences between our two time dependent approaches 3LT$_{topic}$, 3LT$_{tag}$ and 3L at the same $\alpha$ levels.}
  \label{tab:mrr_map_norm}
\end{table*}

In this section we present the evaluation of our two novel algorithms in two steps that correspond to our two research questions. In step 1, we compared the three 3Layers approaches (3L, 3LT$_{topic}$ and 3LT$_{tag}$) with one another to examine our first research question of whether recency has a differential effect on topics and tags. Referring to our empirical analysis in Section \ref{sec:recency}, 3LT$_{tag}$ should yield more accurate predictions than 3LT$_{topic}$ and 3L. 

The results in Table~\ref{tab:mrr_map_norm_layers} are well in accordance with this assumption since - independent of the metric ($F_1$@5, MRR and MAP) and the number of LDA topics (100, 500, and 1000) applied - the difference between 3LT$_{tag}$ and 3L appears to be larger than the one between 3LT$_{topic}$ and 3L. Hence, a user's gist traces (LDA topics) associated with the user's bookmarks) are less prone to ``forgetting'' than a user's verbatim traces (tags associated with the bookmarks). Interestingly, this effect seems to be more strongly pronounced under the narrow folksonomy condition (Flickr) than under the broad folksonomy condition (BibSonomy and CiteULike).

Furthermore, Table~\ref{tab:mrr_map_norm_layers} shows the performance of 3L, 3LT$_{topic}$ and 3LT$_{tag}$ for different numbers of LDA topics (100, 500 and 1000). In general these results reveal that all three approaches provide reasonable results for different levels of topic specialization and that the best accuracy results are reached with 1000 LDA topics\footnote{\textbf{NOTE:} We also performed experiments with more than 1000 LDA topics (e.g., 2000, 3000, ...). However, as also shown by related work (e.g., \cite{krestel2009latent, krestel2010language}) this step did not help in increasing the performance of the LDA-based tag recommenders.}. The $F_1$@5, MRR and MAP values calculated for 1000 topics are also used in Table~\ref{tab:mrr_map_norm} for the second evaluation step that is described in the next paragraph.

In a second step, we contrasted the performance of our approaches, especially 3LT$_{tag}$, with several state-of-the-art algorithms to address our second research question of whether 3L and its two extensions can be implemented in form of effective and efficient tag recommendation mechanisms.
First, Table~\ref{tab:mrr_map_norm} reveals that all personalized recommendation mechanisms clearly outperform the unpersonalized MP approach, which simply takes into account the tag's usage frequency independent of information about a particular user or resource.

Second and more important, 3L and its two extensions (3LT$_{topic}$ and 3LT$_{tag}$) appear to reach higher accuracy estimates than the well-established mechanisms LDA, MP$_{u,r}$, CF, APR, FR, FM and PITF. From this we conclude that predicting tags in form of psychologically plausible steps of calculation that turn a user's gist traces into words yields tag recommendations that correspond well to the user's tagging behavior. 

\begin{figure*}[t!]
   \centering 
   \subfloat[BibSonomy]{ 
      \includegraphics[width=0.30\textwidth]{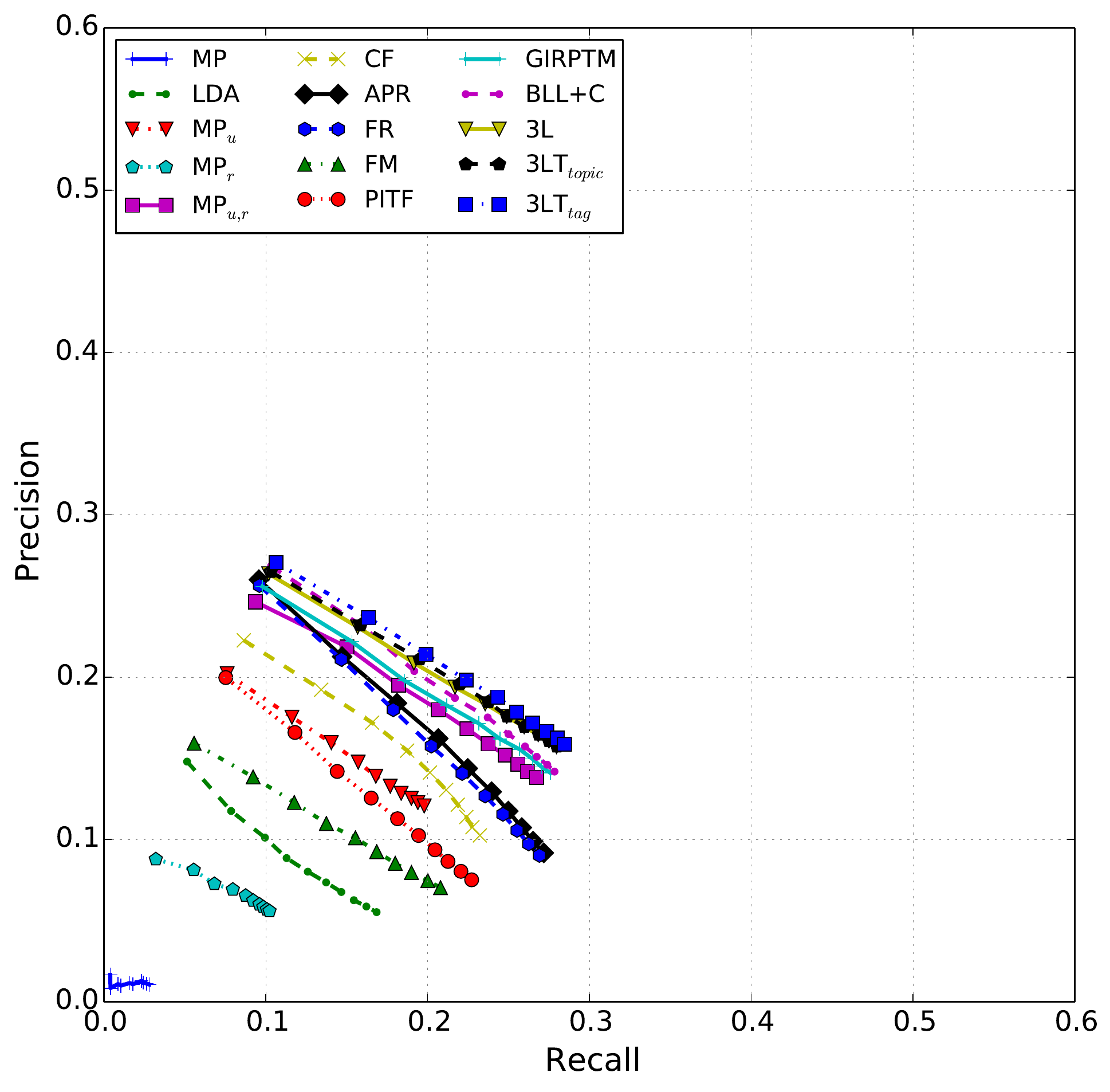}%
   } 
   \subfloat[CiteULike]{ 
      \includegraphics[width=0.30\textwidth]{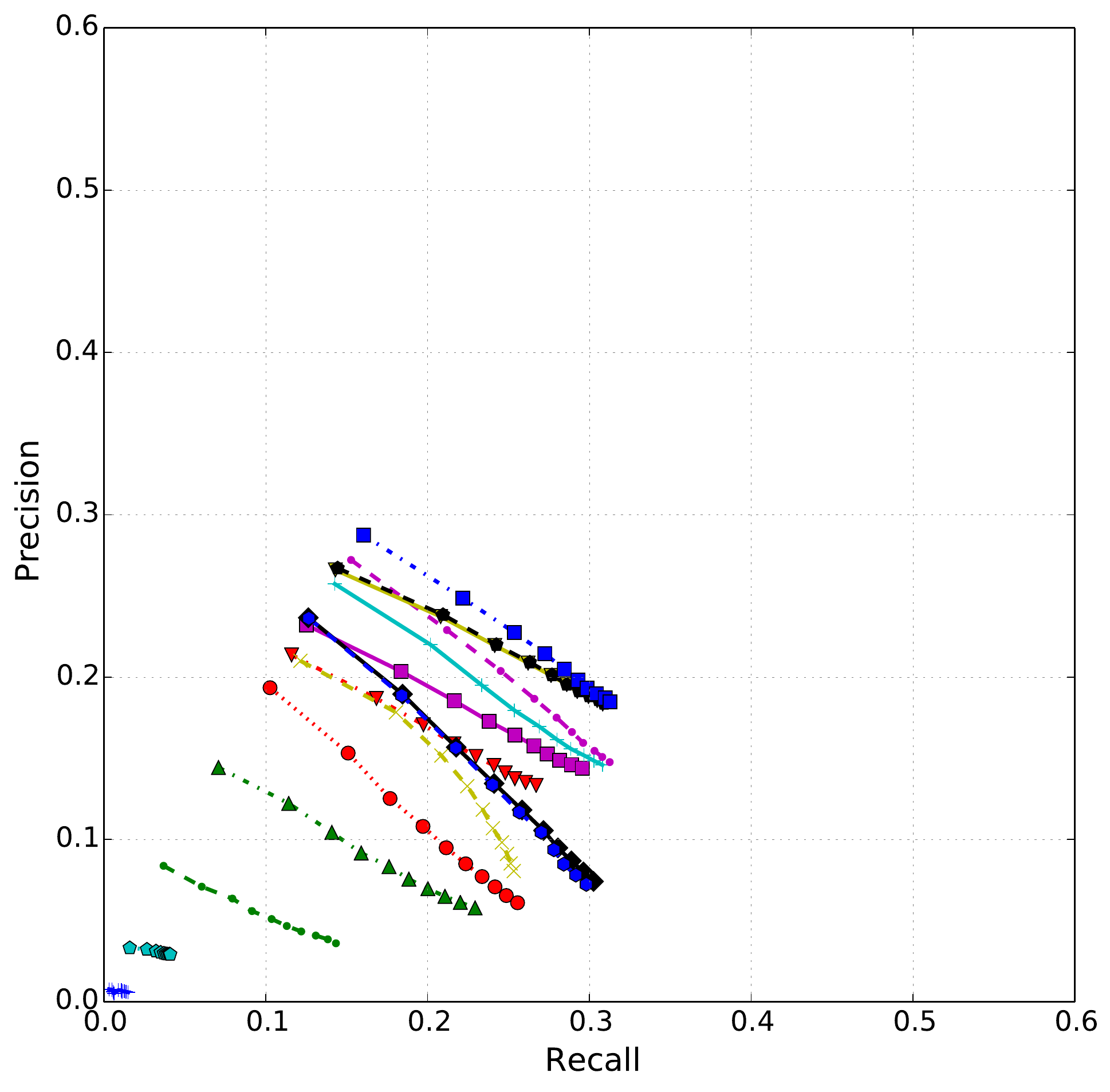}%
   } 
   \subfloat[Flickr]{ 
      \includegraphics[width=0.30\textwidth]{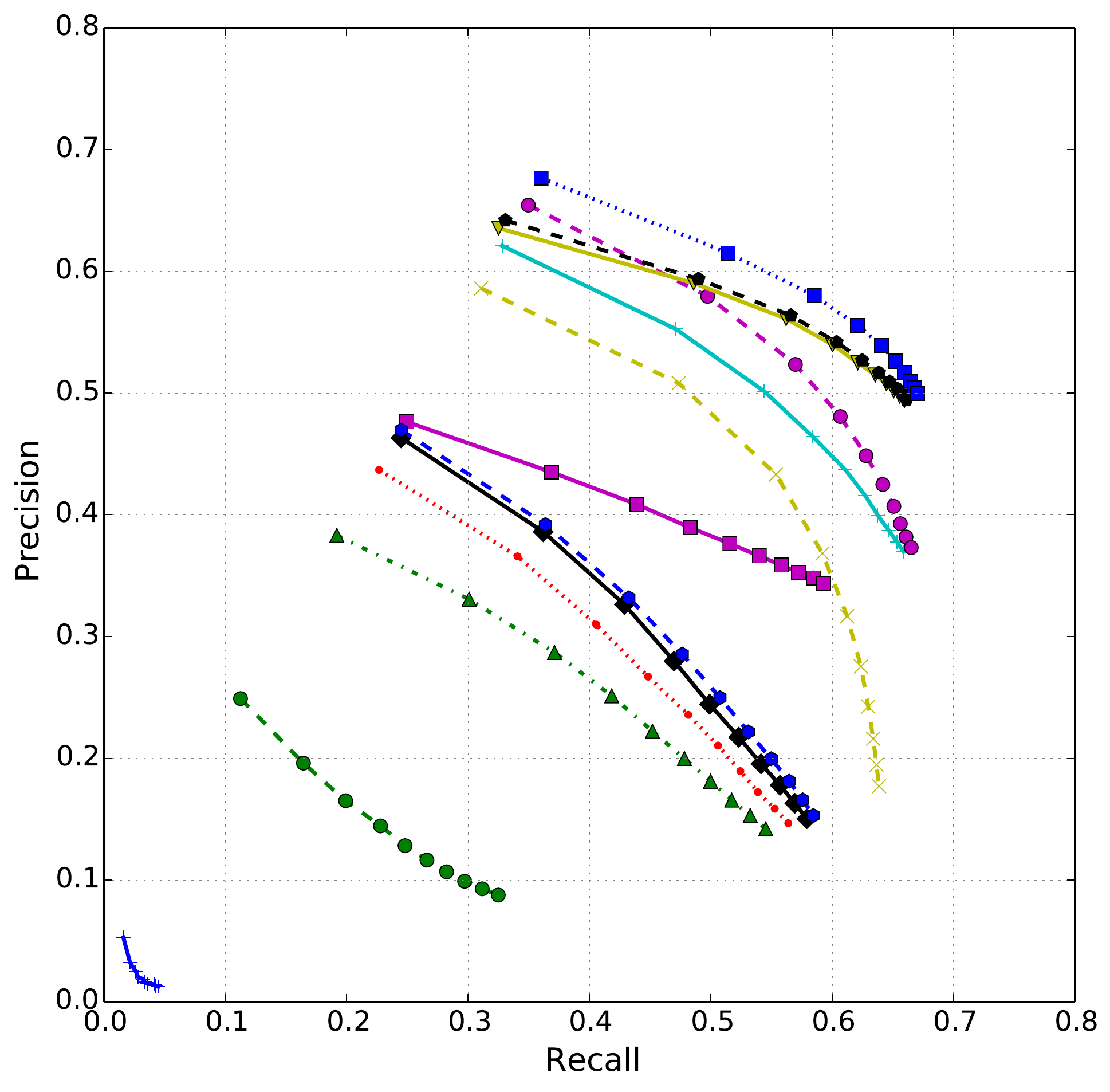}%
   } \\
   \subfloat[BibSonomy (|$B_{min}$|	= 20)]{ 
      \includegraphics[width=0.30\textwidth]{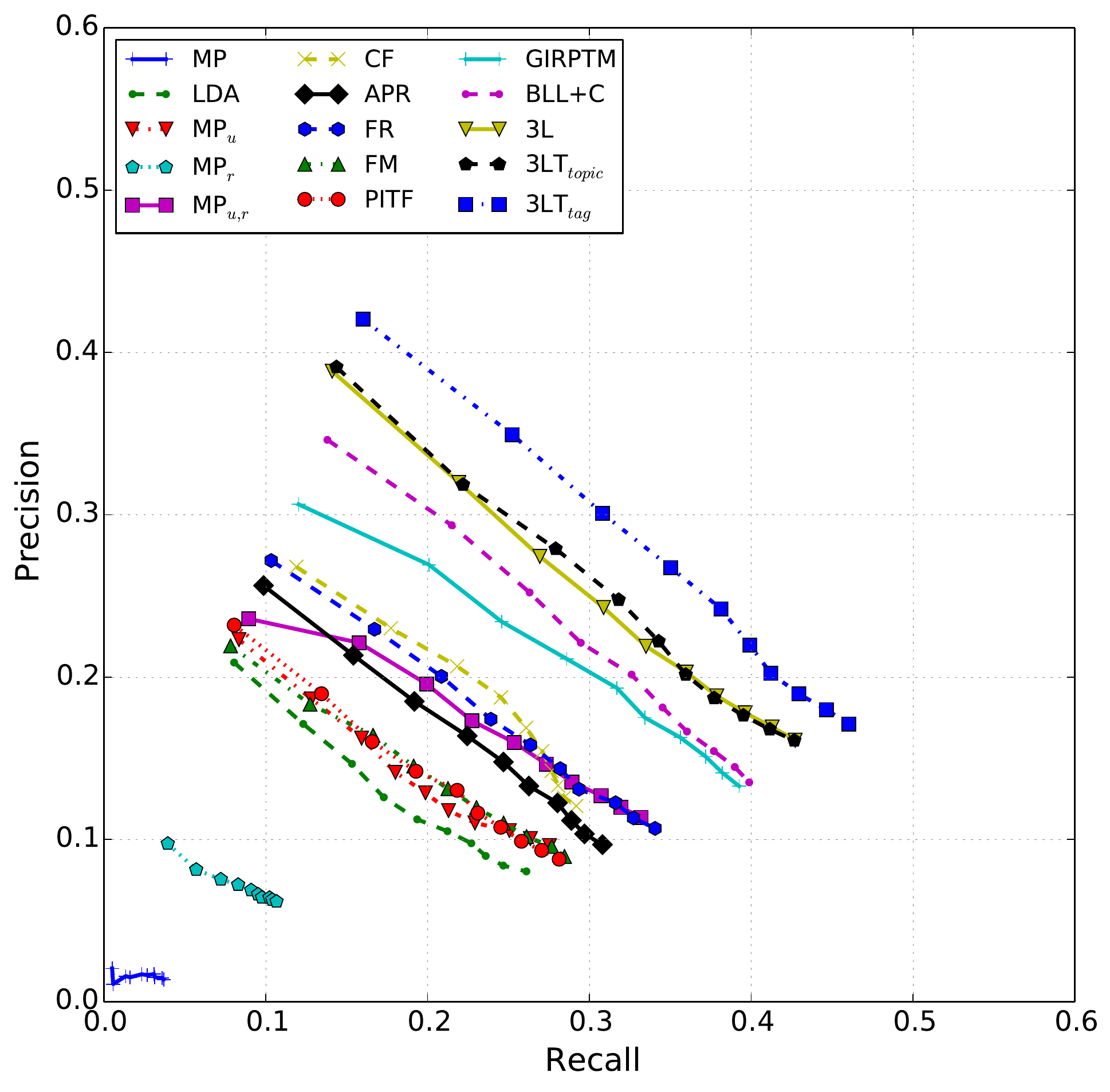}%
   } 
   \subfloat[CiteULike (|$B_{min}$|	= 20)]{ 
      \includegraphics[width=0.30\textwidth]{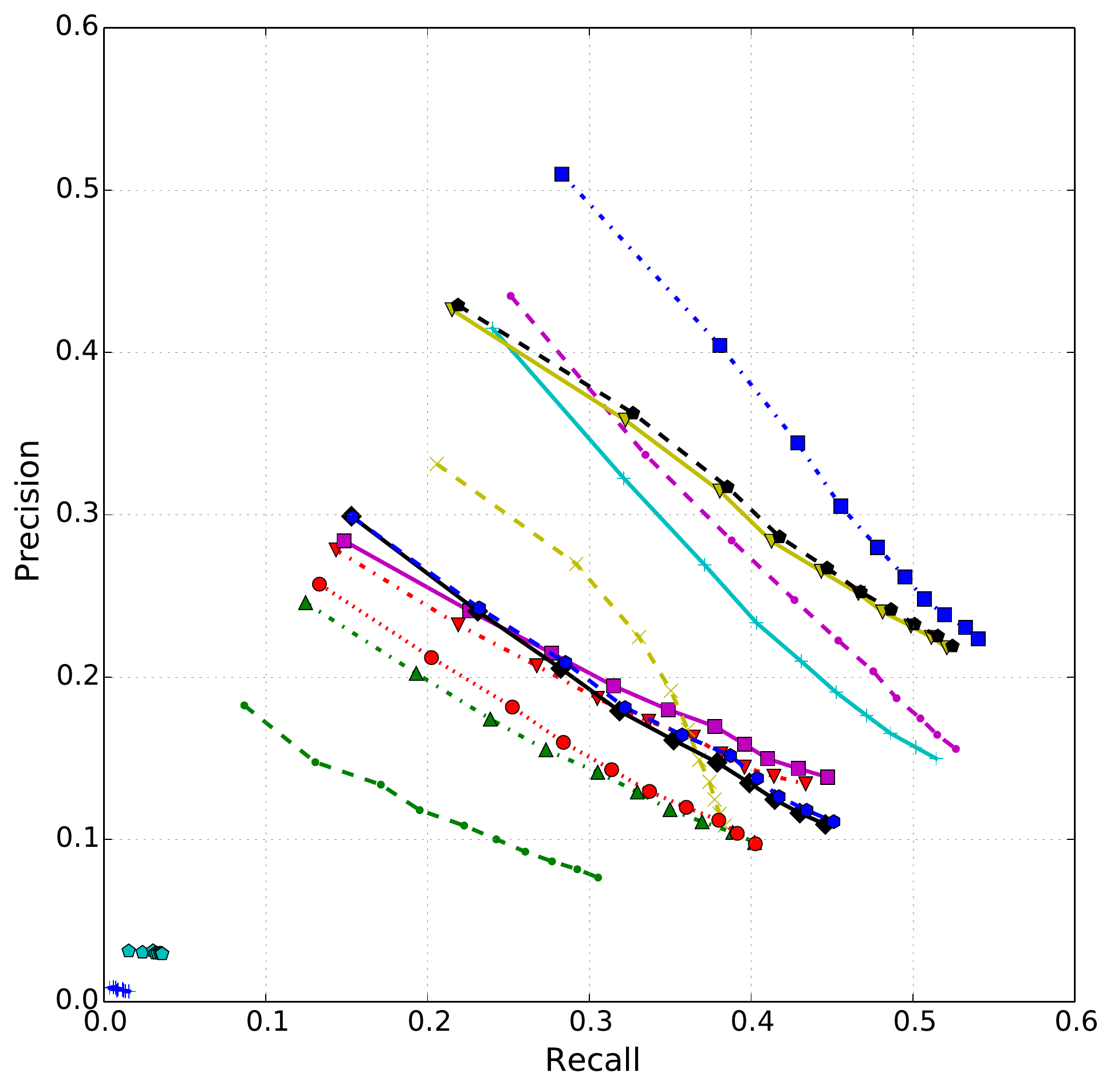}%
   } 
   \subfloat[Flickr (|$B_{min}$| = 20)]{ 
      \includegraphics[width=0.30\textwidth]{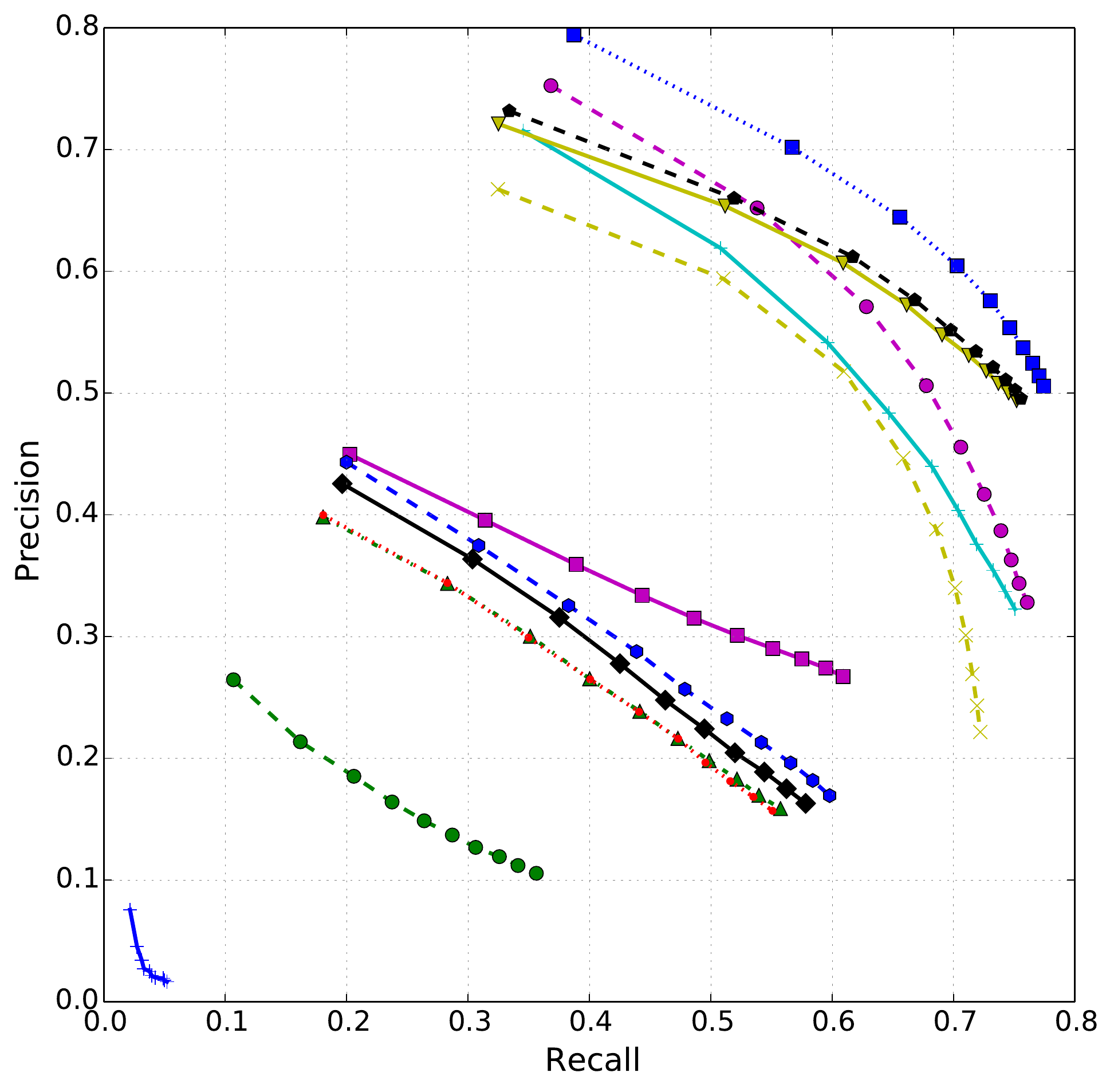}%
   } 
        \caption{Recall/Precision plots for all the users in the datasets (BibSonomy, CiteULike and Flickr) and for users with a minimum number of 20 bookmarks (|$B_{min}$| = 20) showing the performance of the algorithms for 1 - 10 recommended tags (\textit{k}).}
				\label{fig:prec_rec}
\end{figure*}

Third, also the two other time dependent algorithms (GIRPTM and BLL+C) outperform these state-of-the art approaches that do not take the time component into account and in the case of BLL+C also reach higher estimates of accuracy than our 3L approach. However, this relationship between the two mechanisms dramatically changes if we enhance 3L by the recency component at the level of tags. Actually, 3LT$_{tag}$ appears to outperform BLL+C in terms of all three measures and across all three datasets. Finally, as Figure~\ref{fig:prec_rec} shows, a very similar pattern of results becomes apparent if the different approaches are evaluated by plotting recall against precision for $k$ = 1 - 10 recommended tags.

To furthermore proof our assumption that memory processes play an important role in social tagging systems, we also performed an experiment where we looked at users that have bookmarked a minimum of |$B_{min}$| = 20 resources (see also \cite{parra2010improving}). We conducted this experiment by applying a post-filtering method, i.e., recommendations were still calculated on the whole folksonomy graph but accuracy estimates were calculated only on the basis of the filtered user profiles (= 780 users in the case of BibSonomy, 1,757 in the case of CiteULike and 4,420 for Flickr). The results of the experiment are also shown in Table~\ref{tab:mrr_map_norm}. We observe that in general the accuracy estimates of all algorithms are increasing. Furthermore, we can see that the difference between 3LT$_{tag}$ and the other algorithms (including BLL+C)  gets substantially larger the more user ``memory'' (history) is used. These differences between 3LT$_{tag}$ and BLL+C as well as between 3LT$_{tag}$ and 3L proved to be statistically significant based on a Wilcoxon Rank Sum test across all accuracy metrics ($F_1$@5, MRR and MAP) and all three datasets (see Table~\ref{tab:mrr_map_norm}). 


\section{Related Work} \label{sec:relwork}
In contrast to this study, previous research on tag recommender systems has taken a more pragmatist stance, typically
ignoring cognitive psychological models that can help in explaining how people tag (as it was shown in this work).
To date, the two following approaches have been established -- 
folksonomy-based and content-based tag recommender approaches \cite{lipczak2012hybrid}. In our work we focus on folksonomy-based approaches.

The probably most prominent work in this context is the work of Hotho et al. \cite{hotho2006information} who introduced an algorithm called FolkRank (FR)
that has established itself as the most prominent benchmarking tag recommender approach over the past few years.
Subsequent and other popular works in this context are the studies of J\"{a}schke et al. \cite{jaschke2007tag} or Hamouda \& Wanas \cite{hamouda2011put}
who introduced a set of Collaborative Filtering (CF) approaches for the problem of recommending tags to the user in a personalized manner.
More recent and to some extent also well-know works are e.g., the studies of Rendle et al. \cite{Rendle2010}, Krestel et al. \cite{krestel2009latent}, Rawashdeh et al. \cite{rawashdeh2012folksonomy}, Yin et al. \cite{yin2011exploiting} or Zhang et al. \cite{zhang2012integrating}
who introduce a factorization model, a semantic model (based on LDA), a link prediction model or a time-based model to recommend tags to users (see Section \ref{sec:algos}).

Although the latter mentioned approaches perform more or less well in accurately predicting the users tags, all of them
ignore well-established and long standing research from cognitive psychology on how
humans process information. To bridge this gap we have recently introduced two simple and psychological plausible methods \cite{paul2013,domi2014} (= 3L and BLL+C) that are able (with limitations) to explain memory processes in social tagging systems. 
Based on these studies and new observations made in the current work, we were able to present a novel time-based tag recommender algorithm (= 3LT$_{tag}$) in the end that significantly outperforms the state-of-the-art.

\section{Discussion and Conclusion} \label{sec:disc_con}
In this study we have provided empirical evidence for an interaction between the level of knowledge representation (semantic vs. lexical) and time-based forgetting in the context of social tagging. Based on the analysis of three large-scale tagging datasets we conclude that - as expected - the gist traces of a user's personomy (the combination of LDA topics associated with the bookmarks) are more stable over time than the verbatim traces (the combination of associated tags). This pattern of results is well in accordance with research on human memory (e.g., \cite{brainerd2010recollective}) suggesting that while people tend to forget surface details they keep quite robust memory traces of the general meaning underlying the experiences of the past (e.g., the meaning of read words). The interaction effect suggests that it is worthwhile to differentiate both time-based forgetting as well as level of knowledge representation in social tagging research.

Furthermore, the differential affect of forgetting on the two levels of processing has further substantiated the differences between tagging behavior on a semantic level of gist traces and a lexical level of verbatim traces \cite{Seitlinger2012}. This in turn is in line with cognitive research on social tagging (e.g., \cite{fu2012collaborative}) that suggests to consider a latent, semantic level (e.g., modeled in form of LDA topics) when trying to understand the variance in the statistical patterns on the manifest level of users' tagging behavior.

Finally, we have gathered further evidence for our assumption that interactive systems can be improved by basing them on a thorough understanding of how humans process information. We note in particular that integrating two fundamental principles of human information processing, time-based forgetting and differentiating into semantic and lexical processing, enhances the accuracy of tag predictions as compared to a situation when only one of the principles is considered.
3L, that is enhanced by forgetting on the lexical level (3LT$_{tag}$), outperforms both the traditional 3L, as well as other well-established algorithms, such as CF, APR, FR, FM, PITF and the time-based GIRPTM. Furthermore, 3LT$_{tag}$ also reaches higher levels of accuracy than BLL+C, the to-date leading time-based tag recommender approach, that was introduced in our previous work \cite{domi2014}.

In future work, we plan to include our algorithms in a real online social tagging system (e.g., BibSonomy). Only in such setting is it possible to test the recommendation performance by looking at user acceptance. 
Because our approach is theory-driven, it is rather straightforward to transfer it to recommendations in other interactive systems and Web paradigms where semantic and lexical processing play a role (such as, for example, in Web curation). Generalization to other paradigms is another important benefit of driving recommender systems research by an understanding of human information processing on the Web.

\textbf{Acknowledgments:}
The authors would like to thank Andreas Hotho and Denis Parra for many valuable comments on this work. This study is supported by the Know-Center, the EU funded project Learning Layers (Grant Agreement 318209) and the Austrian Science Fund (FWF): P 25593-G22.

\balance
\bibliographystyle{abbrv}

\bibliography{umap2014}

\end{document}